# A Q-Q plot aids interpretation of the False Discovery Rate

NICHOLAS. W. GALWEY

*Research Statistics Group, GlaxoSmithKline Medicines Research Centre,
Gunnels Wood Road, Stevenage, Hertfordshire, SG1 2NY, UK*
nicholas.w.galwey@gsk.com

SUMMARY

A method is demonstrated for representing the false discovery rate (FDR) in a set of *p*-values on a quantile-quantile (Q-Q) plot of the *p*-values. Recognition of this connection between the FDR and the Q-Q plot facilitates both understanding of the meaning of the FDR, and interpretation of the FDR in a particular data set.

## 1. INTRODUCTION

Quantile-quantile (Q-Q) plots and False Discovery Rates (FDRs) are both regularly used in the interpretation of data analyses that produce large sets of *p*-values, such as high-throughput screens of multiple response variables (e.g. gene expression values potentially influenced by a single cause) (Shedden *and others* (2005)) or multiple explanatory variables (e.g. genetic variants potentially associated with a single response) (The Wellcome Trust Case Control Consortium (2007) and Kathiresan *and others* (2009)). However, there is an important connection between these two statistical tools that does not seem to be widely known. A Q-Q plot is valuable both for understanding the meaning of the FDR, and for interpreting the FDRs obtained from a particular set of *m* significance tests.

## 2. METHOD AND INTERPRETATION

The Q-Q plot is produced by plotting the *m* *p*-values, sorted into ascending order, against the corresponding expected quantiles if the null hypothesis ($H_0$) is true for all tests, both variables being transformed to the $-\log_{10}()$ scale. A specified FDR (*q*) can then be represented by a line parallel to the $H_0$ line with slope = 1 passing through the origin (Fig. 1*a*: simulated data). The intercept of the FDR line is $-\log_{10}(q)$, and the largest *p*-value to be declared significant, according to the step-up procedure of Benjamini and Hochberg (1995), is the first one above the line, reading from the left. All *p*-values to the right of this are declared significant *at the specified FDR, whether or not they are above the line*. This representation makes clear the relationship between the *p*-values declared significant and the full set – and one important message that it conveys is that, strictly speaking, the FDR is a property of a *set* of *p*-values, not of any individual value.

The significance threshold ($\alpha$) implied by the choice of $q$ can be obtained by reading off the coordinate of the largest significant $p$-value from the vertical scale, and the proportion of the full set of $p$-values that are declared significant and counted as 'discoveries' can be obtained by reading off the coordinate from the horizontal scale, back-transforming from $-\log_{10}()$ in both cases. Thus in Fig. 1$a$, if $q = 0.3$ is specified (N.B., an FDR that would be considered excessively generous in most applications), then $\alpha = 10^{-0.701} = 0.199$, and a proportion $10^{-0.167} = 0.68$ of the tests (68%, 136 out of $m = 200$) are declared significant.

When the FDR associated with each $p$-value is indicated by the colour of the corresponding plotting symbol, the plot acquires further power as a tool for interpretation, as shown by comparison of Figs 1$a$ and 1$b$. The minimum attainable FDR in Fig. 1$a$ is $q = 0.250$: if a lower value of $q$ is specified, none of the $p$-values are declared significant. This value of $q$ results in a generous significance threshold of $\alpha = 0.089$: 35% of the $p$-values are smaller than this, but the sequence of smaller $p$-values curves back towards the $H_0$ line (the most extreme ones lying below it) so none of them achieves a lower FDR – and this is indicated by their light-red plotting symbol. Fig. 1$b$ (simulated data) represents a data set with a very different pattern, in which the smaller $p$-values lie further and further from the $H_0$ line, and are associated with smaller and smaller FDRs: if the relatively stringent value $q = 0.1$ is specified, 5% of them (10 out of 200) are still declared significant, and these are readily identified by their deep-red plotting symbol.

Patterns like that in Fig. 1$a$ commonly occur in practice, and result in a not-strictly-monotonic relationship: there are regions of the ordered sequence of $p$-values in which the associated FDR does not change. This may be puzzling until the reason is elucidated by the corresponding Q-Q plot. Such patterns are often an indication of positive correlations among the significance tests, i.e. that they are testing related hypotheses, perhaps due to correlations among the explanatory or response variables. The effective number of tests is then less than $m$, the small $p$-values are not as small as would be expected on the basis of $m$ independent tests, and the FDR is conservative (Benjamini and Yekutielim (2001), Theorem 1.2).

## 3. CONCLUSION

Thus a Q-Q plot with the conventions and annotations described here can be of great value both to the theoretician seeking to understand the FDR, and to the practitioner seeking to explain the signals in the data to an audience.

## ACKNOWLEDGEMENTS

The author is grateful to the members of GlaxoSmithKline's Research Statistics Group for helpful comments on the presentation of this method.

*Conflicts of interest:* None declared.

## FUNDING

## REFERENCES

Shedden, K., Chen, W., Kuick, R. *and others.* (2005) Comparison of seven methods for producing Affymetrix expression scores based on False Discovery Rates in disease profiling data. *BMC Bioinformatics* **6**, 26. https://doi.org/10.1186/1471-2105-6-26

The Wellcome Trust Case Control Consortium. (2007) Genome-wide association study of 14,000 cases of seven common diseases and 3,000 shared controls. *Nature* **447**, 661–678. https://doi.org/10.1038

Kathiresan, S., Willer, C., Peloso, G. *and others.* (2009) Common variants at 30 loci contribute to polygenic dyslipidemia. *Nature Genetics* **41**, 56–65. https://doi.org/10.1038/ng.291

Benjamini, Y. and Hochberg, Y. (1995) Controlling the false discovery rate: a practical and powerful approach to multiple testing. *Journal of the Royal Statistical Society B* **57**, 289-300.

Benjamini, Y. and Yekutieli, D. (2001) The control of the false discovery rate in multiple testing under dependency. *The Annals of Statistics* **29**, 1165–1188.

Fig. 1. Quantile-quantile (Q-Q) plots with conventions and annotations to indicate the False Discovery Rate (FDR)

*a*) for a set of hypotheses in which the minimum FDR is attained at an intermediate value of $p_{obs}$

*b*) for a set of hypotheses in which the minimum FDR is attained at the smallest value of $p_{obs}$

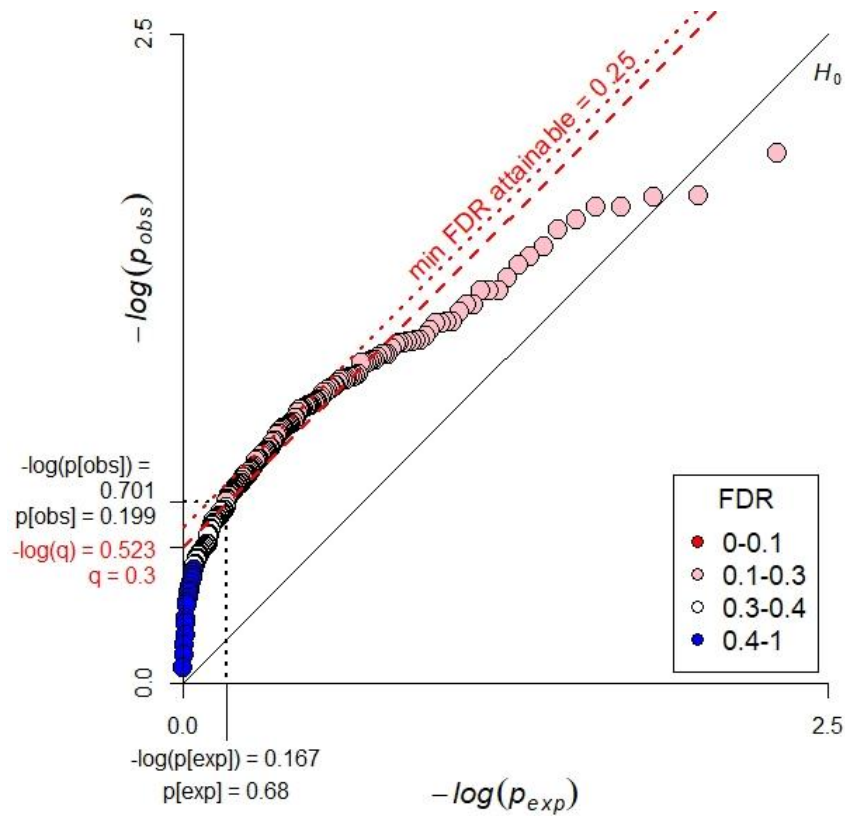
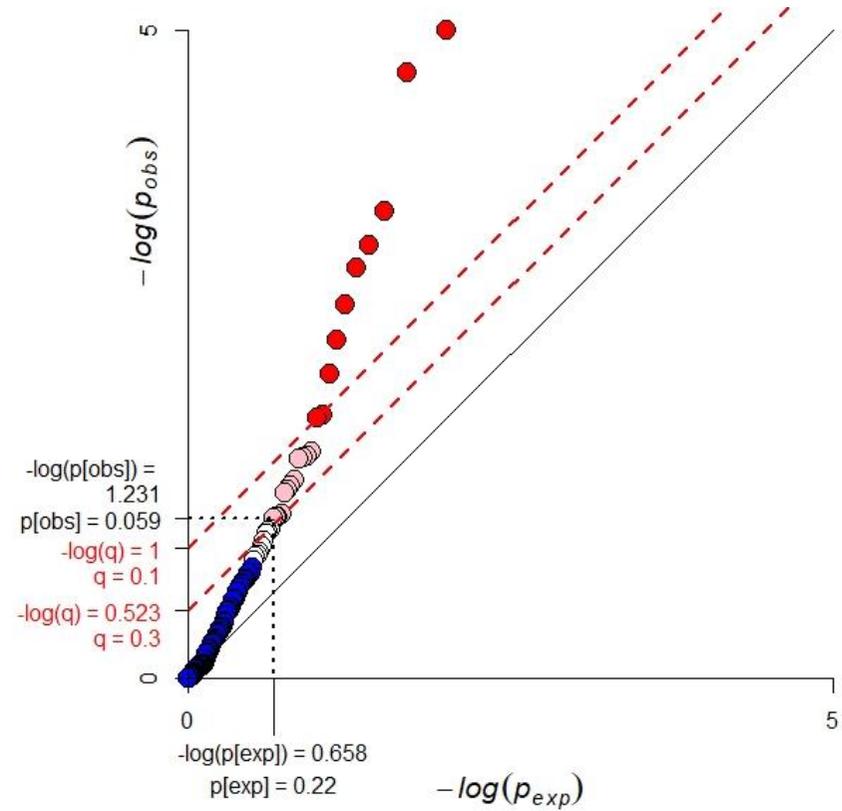